\documentclass[aps,pra,preprint,showpacs,showkeys]{revtex4-1}
\usepackage{amssymb,bm}
\usepackage{graphicx}
\usepackage{amsmath}
\usepackage{epstopdf}
\usepackage{float}
\allowdisplaybreaks
\renewcommand{\Re}{\,\mathrm{Re}}
\renewcommand{\Im}{\,\mathrm{Im}}
\begin{document}

\title{Small-angle scattering  and quasiclassical approximation beyond leading order}

\author{P. A. Krachkov}\email{peter\_phys@mail.ru}
\author{R. N. Lee}\email{R.N.Lee@inp.nsk.su}
\author{A. I. Milstein}\email{A.I.Milstein@inp.nsk.su}
\affiliation{Budker Institute of Nuclear Physics, 630090 Novosibirsk, Russia}

\date{\today}

\begin{abstract}

%BEGIN_FOLD
In the present paper we examine the accuracy of the quasiclassical approach on the example of small-angle electron elastic scattering. Using the quasiclassical approach, we derive the differential cross section and the Sherman function for arbitrary localized potential at high energy. These results are exact in the atomic charge number and correspond to the leading and the next-to-leading high-energy small-angle asymptotics for the scattering amplitude. Using the small-angle expansion of the exact amplitude of electron elastic scattering in the Coulomb field, we derive the cross section and the Sherman function with a relative accuracy $\theta^2$ and $\theta^1$, respectively ($\theta$ is the scattering angle). We show that the correction of relative order $\theta^2$ to the cross section, as well as that of relative order $\theta^1$ to the Sherman function, originates not only from the contribution of large angular momenta $l\gg 1$, but also from that of $l\sim 1$. This means that, in general, it is not possible to go beyond the accuracy of the next-to-leading quasiclassical approximation without taking into account the non-quasiclassical terms.
%END_FOLD

\end{abstract}

\pacs{ 12.20.Ds, 32.80.-t}

\maketitle

\section{Introduction}

In the high-energy QED processes in the atomic field, the characteristic angles $\theta$ between the momenta of final and initial particles are small. Therefore, the main contribution to the amplitudes of the processes is given by the large angular momenta $l\sim \varepsilon\rho \sim \varepsilon/\Delta  \sim 1/\theta$, where $\varepsilon$, $\rho$, and $\Delta$ are the characteristic energy, impact parameter, and momentum transfer, respectively ($\hbar=c=1$). The quasiclassical approach provides a systematic method to account for the contribution of large angular momenta. It was successfully used for the description of numerous processes such as charged particle bremsstrahlung, pair photoproduction, Delbr\"uck scattering, photon splitting, and others \cite{BethMax1954,DavBeMa1954,OlsMaWe1957,OlseMax1959,Milstein:1994zz,LeMaMST2003,KrachkovLeeMilstein2014,KrachkovLeeMilstein2015}. The accurate description of such QED processes is important for the data analysis in modern detectors of elementary particles. The quasiclassical approach allows one to obtain the results for the amplitudes not only in the leading quasiclassical approximation but also with the first quasiclassical correction taken into account \cite{LeeMilsteinStrakhovenko2000,LeeMiSt2004,LeMiStS2005,LeeMil2009b,Lee2011a,KrachkovMilstein2015}.

A natural question arises: how far can we advance in increasing accuracy within the quasiclassical framework? In this paper we examine this question by considering the process of high-energy small-angle scattering of polarized electrons in the atomic field. The general form of this cross section reads (see, e.g., Ref. \cite{uberall2012electron})
\begin{gather}\label{eq:dSigma}
\frac{d\sigma}{d\Omega}= \frac12\frac{d\sigma_0}{d\Omega}\left[1+S\,\bm{\xi}\cdot(\bm\zeta_1+\bm\zeta_2)+T^{ij}\zeta_1^i\zeta_2^j\right]\,,\quad
\bm\xi=\frac{\bm p\times\bm q}{|\bm p\times\bm q|}\,,
\end{gather}
where ${d\sigma_0}/{d\Omega}$ is the differential cross section of unpolarized scattering, $\bm p$ and $\bm q$ are the initial and final electron momenta, respectively, $\bm\zeta_1$ is the polarization vector of the initial electron, $\bm\zeta_2$ is the detected polarization vector of the final electron, $S$ is the so-called Sherman function, and $T^{ij}$ is some tensor. In Section \ref{sec:QC} we use the quasiclassical approach to derive the small-angle expansion of the cross section of electron elastic scattering in arbitrary localized potential. As for the unpolarized cross section ${d\sigma_0}/{d\Omega}$, its leading and subleading terms with respect to the scattering angle $\theta$ are known for a long time \cite{AkhiezerBoldyshevShulga1975a}. They can both be calculated within the quasiclassical framework. We show that the Sherman function $S$ in the leading quasiclassical approximation is proportional to $\theta^2$.  We compare this result with that obtained by means of the expansion with respect to the parameter $Z\alpha$,  \cite{Mott1929,McKinleyFeshbach1948,Dalitz1951,JohnsonWeberMullin1961,GlucLin1964} ($Z$ is the nuclear charge number, $\alpha\approx1/137$ is the fine structure constant). The leading in $Z\alpha$ contribution to the Sherman function is due to the interference between the first and second Born terms in the scattering amplitude. In contrast to the quasiclassical result (proportional to $\theta^2$), it scales as $\theta^3$ at small $\theta$. There is no contradiction between these two results because the expansion of our quasiclassical result with respect to $Z\alpha$ starts with $(Z\alpha)^2$. Therefore, depending on the ratio $Z\alpha/\theta$, the dominant contribution to the  Sherman function is given either by the leading quasiclassical approximation or by the interference of the first two terms of the Born expansion. One could imagine that the terms $O(\theta^3)$ in the function $S$ can be ascribed to the next-to-leading quasiclassical correction and, therefore, they come from the contribution of large angular momenta. However, by considering the case of a pure Coulomb field, we show in Section \ref{sec:Coul} that the account for the angular momenta $l\sim 1$ is indispensable for these terms. Thus, we are driven to the conclusion that, in general, it is not possible to go beyond the accuracy of the next-to-leading quasiclassical approximation without taking into account the non-quasiclassical terms.

\section{Scattering of polarized electrons in the quasiclassical approximation}\label{sec:QC}
It is shown in Ref. \cite{Downie2013} that the wave function $ \psi_{\bm p}(\bm r )$ in the arbitrary localized potential $V(r)$ can be written as
\begin{eqnarray}\label{wf}
&&\psi _{\bm p}(\bm r )=[g_0(\bm r,\bm p)-\bm\alpha\cdot\bm g_1(\bm r,\bm p)-\bm\Sigma\cdot\bm g_2(\bm r,\bm p)]u _{\bm p}\,,\nonumber\\
&& u_{\bm p}=\sqrt{\frac{\varepsilon+m}{2\varepsilon}}
\begin{pmatrix}
\phi\\
\dfrac{{\bm \sigma}\cdot {\bm
p}}{\varepsilon+m}\phi
\end{pmatrix}\,,\quad
\end{eqnarray}
where $\phi$ is a spinor, $\bm\alpha=\gamma^0\bm\gamma$, $\bm\Sigma=\gamma^0\gamma^5\bm\gamma$, $m$ is the electron mass, and $\bm\sigma$ are the Pauli matrices. In this section we assume that $m/\varepsilon\ll 1$. In the leading quasiclassical approximation, the explicit forms of the functions $g_0$  and $\bm g_1$,  as well as the first quasiclassical correction to $g_0$, are obtained in Ref.\cite{LeeMilsteinStrakhovenko2000}.
The first quasiclassical correction to $\bm g_1$ and the leading contribution to $\bm g_2$ are derived in Ref.\cite{KrachkovMilstein2015}.
The asymptotic form of the function $ \psi_{\bm p}(\bm r )$ at large distances $r$ reads
\begin{eqnarray}\label{wf1}
\psi _{\bm p}(\bm r )\approx e^{i\bm p\cdot \bm r}u _{\bm p}+\frac{e^{ip r} }{r}[G_0-\bm\alpha\cdot\bm G_1-\bm\Sigma\cdot\bm G_2]\,u _{\bm p}\,.
\end{eqnarray}
The functions $G_0$, $\bm G_1$, and $\bm G_2$ can be easily  obtained from the expressions for  $g_0$, $\bm g_1$, and $\bm g_2$ in Ref.\cite{KrachkovMilstein2015}:
\begin{align}\label{G012}
G_0&=f_0+\delta f_0\,,\quad \bm G_1=-\frac{\bm \Delta_{\perp}}{2\varepsilon}[f_0+\delta f_0+\delta f_1]\ ,\qquad \bm G_2=i\frac{[\bm q\times\bm p]}{2\varepsilon^2}\delta f_1 \ ,
\end{align}
where
\begin{align}\label{eq:f0etc}
f_0&=-\frac{i\varepsilon}{2\pi}\int d\bm\rho\, e^{-i\bm\Delta_{\perp}\cdot\bm\rho}\left[e^{-i\chi(\bm\rho)}-1\right]\ ,\nonumber\\
\delta f_0&=-\frac{1}{4\pi}\int d\bm\rho\, e^{-i\bm\Delta_{\perp}\cdot\bm\rho-i\chi(\bm\rho)}
 \rho \frac{\partial}{\partial \rho} \int\limits_{-\infty}^\infty dx V^2(r_x)\,,\nonumber\\
\delta f_1&=\frac{i}{4\pi\Delta_{\perp}^2}\int d\bm\rho \,e^{-i\bm\Delta_{\perp}\cdot\bm\rho-i\chi(\bm\rho)}
\bm \Delta_{\perp}\cdot\frac{\bm\rho}{\rho}\frac{\partial }{\, \partial \rho}\int\limits_{-\infty}^\infty dxV^2(r_x)\ ,\nonumber\\
\chi(\bm\rho)&=\int\limits_{-\infty}^{\infty}dxV(r_x)\,,\quad r_x=\sqrt{x^2+\rho^2}\,.
\end{align}
Here $\bm\Delta=\bm q-\bm p$, $\bm q=p\bm r/r$, $\bm\rho$ is a two-dimensional vector perpendicular to the initial momentum $\bm p$, and the notation $\bm X_\perp=\bm X-(\bm X\cdot\bm n_p)\bm n_p$ is used for any vector $\bm X$, $\bm n_p=\bm p/p$. For small scattering angle $\theta\ll1$, we have $\delta f_0\sim \delta f_1\sim \theta f_0$. Taking this relation into account, we obtain  the following expressions for $\frac{d\sigma_0}{d\Omega}$, $T^{ij}$, and $S$ in Eq. \eqref{eq:dSigma}
\begin{align}\label{eq:sec}
&\frac{d\sigma_0}{d\Omega}=|f_0|^2\left[1+2\Re \frac{\delta f_0}{f_0}\right]\,,\\
&T^{ij}=\delta^{ij}+\theta\epsilon^{ijk}\xi^k\,,\nonumber\\
\label{eq:Sherman}
&S=-\frac{m\theta}{\varepsilon}\Im\frac{\delta f_1}{f_0}\,.
\end{align}
In Eqs.\eqref{eq:sec} and \eqref{eq:Sherman} we  keep only the leading and the next-to-leading terms with respect to $\theta$ in $d\sigma_0/d\Omega$ and $T^{ij}$, and the leading term in the function $S$. The form of $T^{ij}$ is a simple consequence of helicity conservation in ultrarelativistic scattering. The expression for $d\sigma_0/d\Omega$ coincides with that obtained in the eikonal approximation \cite{AkhiezerBoldyshevShulga1975a}.
Note that $f_0\to -f_0^*$, $\delta f_0\to \delta f_0^*$, and  $\delta f_1\to \delta f_1^*$ at the replacement $V\to -V$ as it simply follows from Eq. \eqref{eq:f0etc}. Therefore,  the quasiclassical result for the Sherman function $S$, Eq. \eqref{eq:Sherman}, is invariant with respect to the replacement $V\to -V$. In contrast, the term $2\Re(\delta f_0/f_0)$ in ${d\sigma_0}/{d\Omega}$ in Eq. \eqref{eq:sec}  results in the charge asymmetry in scattering, i.e., in the difference between the scattering cross sections of electron and positron, see, e.g., Ref. \cite{uberall2012electron}. Similarly, the account for the first quasiclassical correction leads to the charge asymmetry in lepton pair photoroduction and bremsstrahlung in an atomic field \cite{Lee2011a,Downie2013,KrachkovMilstein2015}.

Let us specialize Eqs.\eqref{eq:sec} and \eqref{eq:Sherman} to the case of a Coulomb field. Substituting $V(r)=-Z\alpha/r$ in Eq. \eqref{eq:f0etc}, we have
\begin{gather}
f_0=\frac{2\eta}{\varepsilon\theta^{2-2i\eta}}\frac{\Gamma(1-i\eta)}{\Gamma(1+i\eta)}\, ,\nonumber\\
\frac{\delta f_0}{f_0}=\frac{1}{4}\pi\theta\eta h(\eta)\, ,\quad
\frac{\delta f_1}{f_0}=-\frac{\pi\theta\eta h(\eta)}{4(1+2i\eta)}\,,\nonumber\\
h(\eta)=\frac{\Gamma(1+i\eta)\Gamma(1/2-i\eta)}{\Gamma(1-i\eta)\Gamma(1/2+i\eta)}\,,\label{CGdelG}
\end{gather}
where $\eta=Z\alpha$ and $\Gamma(x)$ is the Euler $\Gamma$ function. Then, from  Eqs.\eqref{eq:sec} and \eqref{eq:Sherman} we obtain
\begin{align}\label{eq:CSC}
\frac{d\sigma_0}{d\Omega}&=\frac{4\eta^2}{\varepsilon^2\theta^4}\Bigr[1+\frac{\pi\theta\eta}{2}\Re\, h(\eta)\Bigr]\, ,\\
\label{eq:SFC}
S&=\frac{\pi m\eta\theta^2}{4 \varepsilon}\Im\frac{h(\eta)}{1+2i\eta}\,.
\end{align}
The remarkable observation concerning the obtained Sherman function \eqref{eq:SFC} is that it scales as $\theta^2$ while the celebrated Mott result \cite{Mott1929} for the leading in $\eta$ contribution to $S$ scales as $\theta^3\ln \theta$. There is no contradiction because the expansion of \eqref{eq:SFC} in $\eta$ starts with $\eta^2$, while the Mott result is proportional to $\eta$. Thus, the Mott result is not applicable if $\theta\lesssim \eta$. In the next section we obtain the result \eqref{eq:SFC}, along with smaller corrections with respect to $\theta$, by expanding the exact Coulomb scattering amplitude represented as a sum of partial waves. We show that the Mott result is recovered in the order $\theta^3$, as it should be.

\begin{figure}[H]
	\centering
	\includegraphics[width=0.7\linewidth]{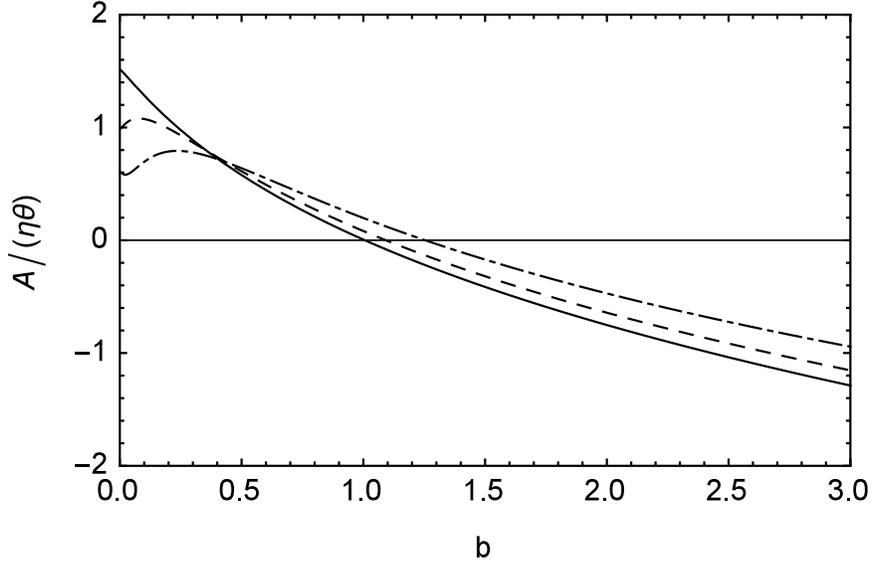}
	\caption{The asymmetry $A$, Eq. \eqref{eq:Afns}, in units $\eta\theta$ as a function of $b=\theta \varepsilon R$ for $\eta=0.1$ (solid curve), $\eta=0.4$ (dashed curve), and $\eta=0.7$ (dash-dotted curve).}
	\label{fig:AoverA0}
\end{figure}

\begin{figure}[H]
\centering
\includegraphics[width=0.7\linewidth]{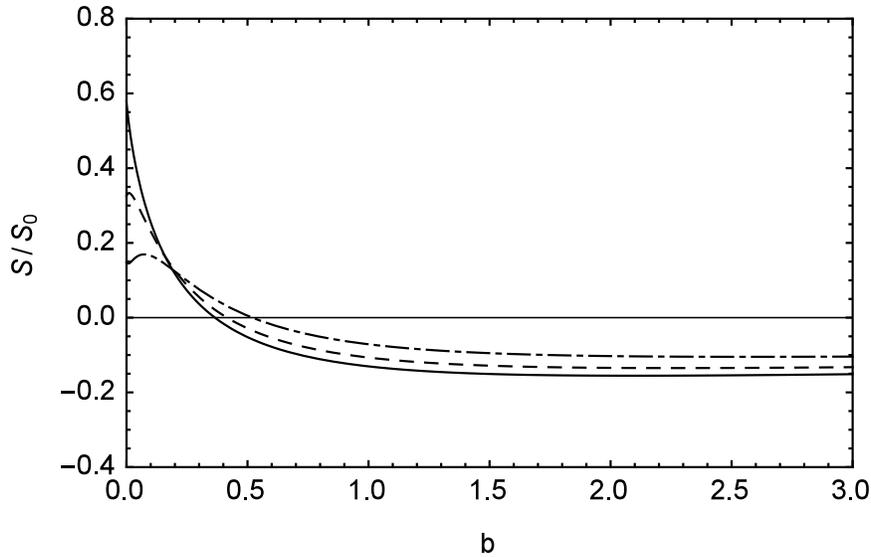}
\caption{The Sherman function $S$, Eq. \eqref{eq:Sfns}, in units $S_0= m \eta^2 \theta^2 /\varepsilon$ as a function of $b=\theta \varepsilon R$ for $\eta=0.1$ (solid curve), $\eta=0.4$ (dashed curve), and $\eta=0.7$ (dash-dotted curve).}
\label{fig:SoverS0}
\end{figure}

Let us now qualitatively discuss the influence of  the finite nuclear size on the cross section ${d\sigma_0}/{d\Omega}$ and the Sherman function $S$. We use the model potential
\begin{equation}
V(r)=-\frac{\eta}{\sqrt{r^2+R^2}}\,,
\end{equation}
where $R$ is the characteristic nuclear size. For this potential we take all integrals in Eq. \eqref{G012} and obtain
\begin{align}\label{eq:CSfns}
\frac{d\sigma_0}{d\Omega}&=\frac{4\eta^2}{\varepsilon^2\theta^4}\left|\frac{b K_{1-i \eta}(b)}{ \Gamma \left(1+i\eta\right)}\right|^2(1+A)\,,\\
\label{eq:Afns}
A&=\frac{\pi\eta\theta}{2}\Re\frac{\Gamma(1+i\eta)(2 K_{1/2-i\eta}(b)-b K_{3/2-i\eta}(b))}{\Gamma(3/2+i\eta)\sqrt{2b}K_{1-i\eta}(b)}\,,\\
\label{eq:Sfns}
S&=\frac{ \pi\eta m \theta^2 }{4\varepsilon}\Im\frac{\Gamma (1+i \eta) K_{{1}/{2}-i \eta}(b)}{ \Gamma \left({3}/{2}+i\eta\right)  \sqrt{2b}\,K_{1-i \eta}(b)}\,,
\quad b = \theta \varepsilon R\,,
\end{align}
where $K_\nu(x)$ is the modified Bessel function of the second kind. The quantity $A$ in Eq. \eqref{eq:Afns} is nothing but the charge asymmetry,
\begin{equation}
A=\frac{d\sigma_0(\eta)-d\sigma_0(-\eta)}{d\sigma_0(\eta)+d\sigma_0(-\eta)}\,.
\end{equation}
As it should be, in the limit $b\to 0$ the results \eqref{eq:CSfns} and \eqref{eq:Sfns} coincide with Eqs. \eqref{eq:CSC} and \eqref{eq:SFC}, respectively.  In Fig. \ref{fig:AoverA0} and Fig. \ref{fig:SoverS0}  we plot the asymmetry $A$ and the Sherman function $S$ as the functions of $b$ for a few values of $\eta$. It is seen that both functions strongly depend on $b$ and $\eta$. It is interesting that they both change their signs at $b\sim1$. Presumably, the latter feature takes place also for the commonly used parametrizations of the nuclear potential.

\section{Small-angle expansion of the Coulomb scattering amplitude}\label{sec:Coul}

In this section we investigate the nontrivial interplay between the contributions of large angular momenta $l$ (quasiclassical contribution) and $l\sim 1$ to the cross section and Sherman function for electron elastic scattering in the Coulomb field. Note that, for small angle $\theta$, the main contribution to the scattering amplitude is given by $l\gg 1$ not only in the ultrarelativistic limit, but for arbitrary $\beta=p/\varepsilon$ as well. Therefore, we treat the parameters $\eta=Z\alpha$ and $\nu=Z\alpha/\beta$ as independent ones. We perform small-angle expansion of the amplitude, but do not assume that $\eta\ll 1$, in contrast to the consideration in Ref.\cite{GlucLin1964}.

The elastic scattering amplitude reads (see, e.g., Refs. \cite{berestetskii1982quantum,uberall2012electron}):
\begin{align*}
M_{fi} &=\frac{i}{2p}\phi_{f}^{\dagger}\left[G\left(\theta\right)-\frac{i \eta m}{p}F\left(\theta\right)-i\left(G\left(\theta\right)\tan\frac{\theta}{2}+\frac{i \eta m}{p}F\left(\theta\right)\cot\frac{\theta}{2}\right)\bm{\xi}\bm{\sigma}\right]\phi_{i}\,,
\end{align*}
where $\phi_{i}$ and $\phi_{f}$ are the spinors of the initial and final electron, respectively.
The functions $F\left(\theta\right)$ and $G\left(\theta\right)$ have the form
\begin{equation}
F\left(\theta\right) =-\sum_{l=1}^{\infty}\frac{\Gamma\left(\gamma_{l}-i\nu\right)}{\Gamma\left(\gamma_{l}+i\nu+1\right)}e^{i\pi\left(l-\gamma_{l}\right)}l\left[P_{l}-P_{l-1}\right]\label{eq:Fdef}\,,\quad G\left(\theta\right)=-\cot\frac{\theta}{2}\frac{dF}{d\theta}\,.
\end{equation}
Here $P_{l}=P_{l}\left(\cos\theta\right)$ is the Legendre polynomial, $\gamma_l=\sqrt{l^2-\eta^2}$.

The unpolarized cross section ${d\sigma}/{d\Omega}$
and Sherman function $S\left(\theta\right)$ are readily expressed
in terms of $F\left(\theta\right)$ and $G\left(\theta\right)$:
\begin{align}
\frac{d\sigma_0}{d\Omega} & =\frac{1}{4p^{2}}\left\{ \frac{\left|G\left(\theta\right)\right|^{2}}{\cos^{2}\frac{\theta}{2}}+\frac{\eta ^{2}m^{2}\left|F\left(\theta\right)\right|^{2}}{p^{2}\sin^{2}\frac{\theta}{2}}\right\} \label{eq:sigma_via_FG}\\
S\left(\theta\right) & =\frac{\eta mp\sin\theta\Re FG^{*}}{\left|G\left(\theta\right)\right|^{2}p^{2}\sin^{2}\frac{\theta}{2}+\eta^{2}m^{2}\left|F\left(\theta\right)\right|^{2}\cos^{2}\frac{\theta}{2}}\nonumber
\end{align}

We want to find the expansion of ${d\sigma_0}/{d\Omega}$ and $S$ with respect to $\theta$. The main contribution to the sum in Eq. (\ref{eq:Fdef}) comes from the region of large $l$. Let us write the function $F$ as
\begin{align}\label{eq:Fa}
F&=F_{a}+F_{b}\,,\nonumber\\
F_{a} &=-\sum_{l=1}^{\infty}\frac{\Gamma\left(l-i\nu\right)}{\Gamma\left(l+i\nu+1\right)}l\,T_l \left[P_{l}-P_{l-1}\right]\,,\nonumber\\
F_{b}&=-\sum_{l=1}^{\infty}\Bigg[\frac{\Gamma\left(\gamma_{l}-i\nu\right)}{\Gamma\left(\gamma_{l}+i\nu+1\right)}e^{i\pi\left(l-\gamma_{l}\right)}-\frac{\Gamma\left(l-i\nu\right)}{\Gamma\left(l+i\nu+1\right)}T_l\Bigg] l\left[P_{l}-P_{l-1}\right]\,,\nonumber\\
T_l&=1+\frac{i\pi}{2l}\eta ^{2}+\frac{\eta ^{2}}{2l^{2}}\left(1+2i\nu-\frac{\pi^{2}\eta ^{2}}{4}\right)\,.
\end{align}
The quantity $T_l$ is the expansion of
$\frac{\Gamma\left(\gamma_{l}-i\nu\right)/\Gamma\left(l-i\nu\right)}{\Gamma\left(\gamma_{l}+i\nu+1\right)/\Gamma\left(l+i\nu+1\right)}e^{i\pi\left(l-\gamma_{l}\right)}$
over $1/l$ up to $O\left(1/l^{2}\right)$. The sum in the definition of $F_a$ can be taken analytically at $\theta\ll 1$.
In order to do this we use the integral representation
\[
\frac{\Gamma\left(l-i\nu\right)}{\Gamma\left(l+i\nu+1\right)}=\frac{1}{\Gamma\left(1+2i\nu\right)}\intop_{0}^{\infty}dy\,\frac{y^{2i\nu}}{\left(1+y\right)^{l+i\nu+1}}\,
\]
and take the sum over $l$ using the generating function for the Legendre polynomials. We obtain
\begin{multline}\label{eq:Fa1}
F_{a}\left(\theta\right)=\frac{1}{\Gamma\left(1+2i\nu\right)}\intop_{0}^{\infty}\frac{dy\, y^{2i\nu}}{\left(1+y\right)^{1+i\nu}}\Bigg\{\frac{2s^{2}\left(2+y\right)\left(1+y\right)}{\varrho^{3}}+\frac{i}{2}\pi\eta ^{2}\left(1-\frac{y}{\varrho}\right)\\
+\frac{\eta ^{2}}{2}\left(1+2i\nu-\frac{\pi^{2}\eta ^{2}}{4}\right)\ln\left[\frac{\left(1-s^{2}\right)\left(2s^{2}+y+\varrho\right)}{y+\varrho-2s^{2}\left(1+y\right)}\right]\Bigg\}\,,
\end{multline}
where $s=\sin\frac{\theta}{2}$ and $\varrho=\sqrt{y^{2}+4s^{2}\left(1+y\right)}$.
As it follows from Eq. \eqref{eq:Fa1}, the convenient variable for the small-angle expansion is $s\ll 1$. There are two regions, which contribute to the integral over $y$:
\[
\mathrm{I}.\ y\sim s\ll 1\,,\quad\mathrm{II}.\ y\sim1\,.
\]
The first region provides contributions $\propto s^{n+2i\nu}$ ($n=0,1,\ldots$),
while the second region provides contributions $\propto s^{n}$ ($n=2,3\ldots$).
Calculating the integral with the method of expansion by regions, see, e.g., Ref. \cite{BeneSmi1998}, we arrive at
\begin{align}\label{eq:Fa2}
F_{a}\left(\theta\right) & \approx \frac{\Gamma\left(1-i\nu\right)}{\Gamma\left(1+i\nu\right)}(t_0+t_1+t_2)\,,\nonumber\\
t_0 & =s^{2i\nu}\,,\quad t_1=i\pi\eta ^{2}\frac{s^{1+2i\nu}}{1+2i\nu}h(\nu)\,,\nonumber\\
t_2& =i\frac{s^{2+2i\nu}\eta ^{2}}{2\left(1+i\nu\right)\nu}\left[1+2i\nu-\frac{\pi^{2}\eta ^{2}}{4}\right]-i s^{2}\eta ^{2}\Bigg[\frac{1}{2\nu}+i+\frac{\pi}{2\left(1-2i\nu\right)}-\frac{\pi^{2}\eta ^{2}}{8\nu}\Bigg]\,.
\end{align}
Here $t_0$ and
$t_1$ correspond, respectively,
to the leading quasiclassical approximation and first quasiclassical correction ($|t_0|=1$, $|t_1|\sim \theta^1$). The relative magnitude of $t_2$ is $\theta^2$ and it is tempting to interpret $t_2$ as a second quasiclassical correction. However, this is not true because the magnitude of $t_2$ is the same as that of the individual terms at $l\sim1$ in the sum in Eq. (\ref{eq:Fdef}). It is easy to check that the contribution to $t_2$
proportional to $s^{2+2i\nu}$ remains intact even if the sum over $l$ starts from some $l_{0}\gg 1$, provided that $l_{0}\ll1/s$. Therefore, this contribution is natural
to identify with the second quasiclassical correction.

\begin{figure}[H]
	\centering
	\includegraphics[width=0.7\linewidth]{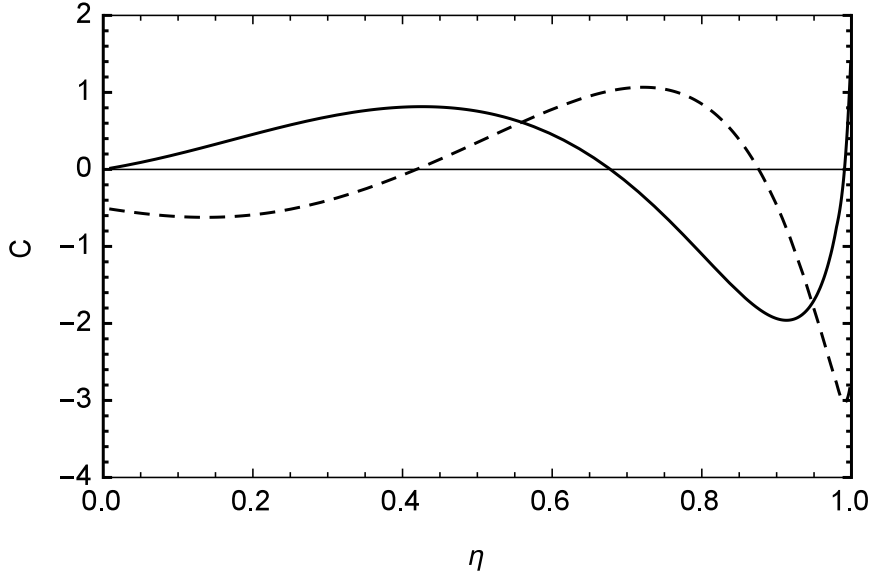}
	\caption{The real (solid curve) and imaginary (dashed curve) parts of $C(\eta,\nu)$, Eq. \eqref{eq:C}, at $\nu=\eta$ as functions of $\eta$.}
	\label{fig:C}
\end{figure}

Let us now consider the function $F_{b}$ in Eq. \eqref{eq:Fa}. The sum over $l$ converges at $l\sim1$, and we can approximate $P_{l}\left(\theta\right)-P_{l-1}\left(\theta\right)$
by $-2ls^{2}$. Since $F_{b}$ in the leading order is proportional
to $s^{2}$, it is natural to sum up $F_{b}$ and the
term in $F_{a}\left(\theta\right)$, Eq. \eqref{eq:Fa2}, proportional
to $s^{2}$. Finally we have
\begin{align}
F & =F_{\text{QC}}+\delta F+O\left(s^{3}\right)\,,\label{eq:Fexp}\\
F_{\text{QC}} & =\frac{\Gamma\left(1-i\nu\right)}{\Gamma\left(1+i\nu\right)}s^{2i\nu}\Bigg[1+\frac{i\pi\eta ^{2}}{1+2i\nu}h\left(\nu\right)s+\frac{i\eta ^{2}}{2\left(1+i\nu\right)\nu}\left(1+2i\nu-\frac{\pi^{2}\eta ^{2}}{4}\right)s^2\Bigg]\,,\nonumber \\
\delta F & =\frac{\Gamma\left(1-i\nu\right)}{\Gamma\left(1+i\nu\right)}C\left( \eta,\nu\right)s^2\,,\nonumber
\end{align}
where
\begin{align}\label{eq:C}
%h\left(\nu\right) & =\frac{\Gamma\left(\frac{1}{2}-i\nu\right)}{\Gamma\left(\frac{1}{2}+i\nu\right)}\frac{\Gamma\left(1+i\nu\right)}{\Gamma\left(1-i\nu\right)}\,,\\
C\left( \eta,\nu\right) & =-i\eta ^{2}\Bigg[\frac{1}{2\nu}+i+\frac{\pi}{2\left(1-2i\nu\right)}-\frac{\pi^{2}\eta ^{2}}{8\nu}\Bigg]\nonumber\\
 & +\frac{\Gamma\left(1+i\nu\right)}{\Gamma\left(1-i\nu\right)}\sum_{l=1}^{\infty}2l^{2}\bigg[\frac{\Gamma\left(\gamma_{l}-i\nu\right)e^{i\pi\left(l-\gamma_{l}\right)}}{\Gamma\left(\gamma_{l}+i\nu+1\right)}-\frac{\Gamma\left(l-i\nu\right)}{\Gamma\left(l+i\nu+1\right)}T_l\bigg]\,,
\end{align}
$T_l$ is defined in Eq. \eqref{eq:Fa}, and $h(\nu)$ is given in Eq. \eqref{CGdelG}. The small-angle expansion of the function $F$ was investigated in Ref. \cite{GlucLin1964} at small $\eta$ and arbitrary $\nu$. Expanding in $\eta$ up to $\eta^4$ under the sum sign in Eq. \eqref{eq:C}  and taking the sum over $l$, we find the agreement with Ref. \cite{GlucLin1964} up to a misprint in Eq. (3.27) of that paper (in the right-hand side of  Eq. (3.27) one should make the replacement $j\to j+1$). The function $C(\eta,\nu)$ strongly depends on the parameters $\eta$ and $\nu$. This statement is illustrated by Fig. \ref{fig:C} where the real and imaginary parts of $C(\eta,\nu)$ at $\nu=\eta$ ($\beta=1$) are shown as functions of $\eta$.

Substituting Eq. (\ref{eq:Fexp}) in Eq. (\ref{eq:sigma_via_FG}), we find
\begin{align}\label{eq:csCoul}
\frac{d\sigma}{d\Omega} & =\frac{\nu ^{2}}{4p^2s^{4}}\bigg[1+\pi \eta\beta\Re h\left(\nu\right)s-2\nu ^{-1}\Im [s^{2i\nu}C^*\left( \eta,\nu\right)]s^{2}\bigg]\,,\\
\label{eq:ShermanCoul}
S\left(\theta\right)= & \frac{ ms^{2}}{\varepsilon\nu}\bigg\{\pi\eta ^{2}\Im\left[\frac{h\left(\nu\right)}{1+2i\nu}\right]+\bigg[\frac{\eta ^{2}}{1+\nu^{2}}\left(1-\frac{3\pi^{2}\eta ^{2}}{4\left(1+4\nu^{2}\right)}\right)\nonumber\\
& -\frac{\pi^{2}\eta ^{4}}{\nu}\Im\left[\frac{h\left(\nu\right)}{1+2i\nu}\right]\Re h\left(\nu\right)-2\Re[\left(1+i\nu\right)s^{2i\nu}C^{*}\left( \eta,\nu\right)]\bigg]s\bigg\}\,.
\end{align}
It is quite remarkable that the second correction to the cross section entirely comes from interference between the quasiclassical and nonquasiclassical terms. Therefore, this correction can not be calculated within the quasiclassical
approach.

We are now in position
to discuss the nontrivial interplay between the small-angle approximation and the small-$\nu$ approximation. Keeping only the leading in $\nu$ terms in the coefficients of the expansion in $s$, we have
\begin{align}\label{eq:dSigmaBorn}
\frac{d\sigma}{d\Omega} & =\frac{\nu ^{2}}{4p^2s^4}(1+  s\pi \eta\beta -s^2\beta^{2})\,,\\
\label{eq:ShermanBorn}
S\left(\theta\right)= & \frac{2\eta m s^2}{\varepsilon}\left[\pi \eta(2\ln2-1)+\beta s \ln s\right]\,.
\end{align}
The cross section \eqref{eq:dSigmaBorn} agrees with the small-angle expansion of the corresponding result in Refs. \cite{McKinleyFeshbach1948,Dalitz1951}. The function $S$, Eq. \eqref{eq:ShermanBorn}, agrees with the small-angle expansion of the Sherman function in Ref. \cite{JohnsonWeberMullin1961}. The term proportional to $s\ln s$ in \eqref{eq:ShermanBorn} corresponds to the celebrated Mott result \cite{Mott1929}.

We see that the relative magnitude of the first and the second corrections with respect to $s$ to the differential cross section is proportional to the ratio $\nu/\theta$ of two small parameters, and this ratio can be smaller or larger than unity. The same phenomenon takes place also in the Sherman function: the ratio of the leading quasiclassical term and the correction is proportional to $\nu/(\theta\ln\theta)$.

\section{Conclusion}\label{conclusion}

In the present paper we have examined the accuracy of the quasiclassical approach when applied to the calculation of the small-angle electron elastic scattering cross section, including the polarization effects. Using the quasiclassical wave function, we have derived the differential cross section with the account of the first correction in $\theta$, Eq. \eqref{eq:sec}, and the Sherman function in the leading order in $\theta$, Eq. \eqref{eq:Sherman}. The results \eqref{eq:sec} and \eqref{eq:Sherman} are valid for ultrarelativistic scattering in the localized central potential of arbitrary strength. In particular, we have investigated the nuclear size effect and found that both the Sherman function and the charge asymmetry (arising from the correction to the cross section) may change their signs in the region where the momentum transfer $\Delta$ is of the order of inverse nuclear radius $R^{-1}$. Using the small-angle expansion of the exact amplitude of electron elastic scattering in the Coulomb field, we have derived the cross section, Eq. \eqref{eq:csCoul}, and the Sherman function, Eq. \eqref{eq:ShermanCoul}, with a relative accuracy $\theta^2$ and $\theta^1$, respectively. The coefficients of the derived expansions in $\theta$ are the exact functions of the parameters $\eta=Z\alpha$ and $\nu=Z\alpha/\beta$. In particular, Eqs. \eqref{eq:csCoul} and \eqref{eq:ShermanCoul} are valid even for $\beta\ll1$. We have shown that the correction of relative order $\theta^2$ to the cross section, as well as that of relative order $\theta$ to the Sherman function, originate not only from the contribution of large angular momenta $l\gg 1$, but also from that of $l\sim 1$. Thus, we are driven to the conclusion that, in general, it is not possible to go beyond the accuracy of the next-to-leading quasiclassical approximation without taking into account the non-quasiclassical terms.

\paragraph*{Acknowledgement.}
This work has been supported by Russian Science Foundation (Project No. 14-50-00080). Partial support of RFBR through Grant No. 14-02-00016 is also acknowledged.
%\bibliography{sherman}
%
\end{document}